\documentclass[letterpaper, 10 pt, conference]{ieeeconf} 
\usepackage{ifpdf}
\usepackage{cite}
\usepackage{amsmath,bm}
\usepackage{amssymb}
\usepackage{algorithmic}
\usepackage{array}
\usepackage{url}
\usepackage{graphicx}
\usepackage{xcolor}
\usepackage{mathtools}
\usepackage{subfigure}
\usepackage{hyperref}

\IEEEoverridecommandlockouts       
\newtheorem{rem}{Remark}
\abovecaptionskip
\belowcaptionskip
\begin{document}
\title{ROS-Based \textbf{M}ulti-\textbf{A}gent \textbf{S}ystems \textbf{CO}ntrol Simulation \textbf{T}estbed (MASCOT)}
\author{Arvind Pandit, Akash Njattuvetty, and Ameer K. Mulla\thanks{Arvind Pandit and Ameer K. Mulla are with the Department of Electrical Engineering, Indian Institute of Technology Dharwad, Karnataka, India. Email: \texttt{arvind.pandit.21@iitdh.ac.in, ameer@iitdh.ac.in}} 
\thanks{Akash Njattuvetty is with the Department of Electrical Engineering, Indian Institute of Technology Madras, Chennai, India. Email: \texttt{akashn2772@gmail.com}}}

\maketitle
\begin{abstract}
This paper presents a simulation testbed developed for testing and demonstration of decentralized control algorithms designed for multi-agent systems. Aimed at bridging a gap between theory and practical deployment of such algorithms, this testbed provides a simulator with multi-agent systems having quadcopter agents. It is used to test the control algorithms designed assuming simple agent dynamics like a single or double integrator. This is based on the fact that under certain assumptions, quadcopter dynamics can be  modeled as a double integrator system. A gazebo simulator with a physics engine as ODE (Open Dynamics Engine) is used for simulating the dynamics of the quadcopter. Robot Operating System is used to develop communication networks and motion control algorithms for the different agents in the simulation testbed. The performance of the test bed is analyzed by implementing linear control laws such as position control, leaderless consensus, leader-follower consensus, and non-linear control law for min-max time consensus. This work is published as an open-source ROS package under MIT license at \url{https://github.com/Avi241/mascot}. A docker image is also developed for easy setup of the system.
\end{abstract}


\section{Introduction}

Distributed control of multi-agent systems has been a very active area of research over the last couple of decades, starting from the asymptotic consensus laws proposed in \cite{asymptotic_control_laws}, various distributed control laws for finite-time consensus \cite{distributed_control_laws_for_finite_time_consensus_double_integrator}, formation control \cite{formation_control}, trajectory tracking \cite{trajectory_tracking}, applications like survey, search and rescue \cite{search_and_rescue}, non-cooperative tasks like pursuit-evasion \cite{pursuit_evasion} etc. The primary reason for the popularity of multi-agent systems is that multi-agent systems are much more capable than individual agents and can achieve complex objectives using simple agents. 
Most of such control algorithms are designed assuming simple dynamics of agents, for example, single integrator, double integrator \cite{modeling_integrator}, non-holonomic Dubin's vehicle model, etc. These agent models are popular mainly because the dynamics of various robots and autonomous vehicles, under a few assumptions, may be simplified to such models.

With the developments in the field of robotics and communication, Unmanned Aerial and Ground Vehicles (UAVs and UGVs) are being used in various applications related to agriculture \cite{multiagent_agriculture}, disaster response \cite{multiagent_disaster}, crowd monitoring, event monitoring, and recording \cite{monitoring} and so on. While UGVs and fixed-wing UAVs are used for certain specific applications, quadcopters are used more widely due to their advantages such as high maneuverability, 6-DoF, easy availability, etc\cite{QuadDynamics}. These advantages make the quadcopters suitable for deploying multi-agent co-operative control algorithms in the testing stage. In most cases, even though multiple quadcopters are used, their control is either centralized or each agent is controlled independently. 

The use of distributed control can potentially increase the capabilities and efficiency of multi-agent systems. However, such control laws are seldom used, primarily because of the problems encountered during deploying the control algorithms on physical systems. Even though it is shown that quadcopter dynamics can be approximated to a double integrator, before deploying the control laws designed for double-integrator on quadcopters \cite{implementation_of_distributed_consensus}, it is essential to ensure that the simplifying assumptions are satisfied by the actual system and the control parameters need to be tuned appropriately. 
Extensive simulations are necessary for appropriate tuning of the controllers before actual deployment on physical systems to ensure that the systems remain stable and the controller will work as desired. A good testing platform with a simple user interface can address this problem. MASCOT is designed to be such a testing platform. In the current version, MASCOT uses quadcopters as agents to simulate multi-agent systems. The lower-level controllers of the quadcopters are tuned so that the user can consider the agents to have double integrator dynamics in the primary directions and test the control algorithms. By adjusting the control parameters of the lower-level controllers, practical scenarios like relaxing the assumptions may be created. 

There are few MATLAB-based quadcopter simulators that are either for the single quadcopter or lack the environmental dynamics and visuals\cite{Matlabdrone2}. MATLAB provides a few toolboxes which help for testing the algorithms but there are limitations of MATLAB such as inter-process communications, physical dynamics, code porting from simulation to hardware, etc. Some authors have demonstrated the control algorithms using realistic simulation environments and /or hardware test-beds.  \cite{implementation_of_distributed_consensus,Implementation_min_max_consensus}. However, these efforts are limited to specific control algorithms and are not easily accessible to other researchers. Moreover, such hardware testbeds come with physical limitations such as a number of agents, space, cost, as well as the potential risk of damage to the vehicles and the surroundings. 

MASCOT is developed using open-source software, the Robot Operating System (ROS) \cite{ros} and Gazebo. Gazebo with its ODE (Open Dynamics Engines) provides a simulation environment with various real-world physical dynamics such as collision, gravity, mass, inertia, wind effects, and many more which are very tedious to configure in MATLAB. Moreover, ROS supports low-level drivers and programming in C++, making it very convenient to directly deploy the simulation codes on physical systems, which are not straightforward in MATLAB. ROS also provides a rich set of open-source packages for different applications which helps researchers to modify and use the code as per their needs. MASCOT retains all of the advantages of ROS while simplifying the user interface by providing simple text files for configuring the agents and control laws for simulations.
While this paper describes the simulations of multi-agent systems with double integrator agents, simulating systems with single integrator agents, non-holonomic agents and heterogeneous systems are equally easy on MASCOT.

The rest of the paper is organized as follows. In Section \ref{sec:prelim}, we discuss the quadcopter dynamics. Section \ref{sec:mascot} describes the structure and features of the designed simulation testbed along with the tools used in the development. In Section \ref{sec: Use Cases}, various use cases of the testbed are discussed with the simulations carried out using MASCOT. We conclude in \ref{sec:conc} with some remarks on the applications and future developments planned for the test-bed.



\section{Preliminaries}\label{sec:prelim}
In this section, first, we describe the preliminaries of the dynamics of a quadcopter and review how a quadcopter can be represented using double integrator dynamics. A few preliminaries used in distributed control of multi-agent systems are also discussed.
\subsection{Frame of References}\label{subsec:for}
\begin{figure}[!ht]
\centering
    \includegraphics[scale=0.40]{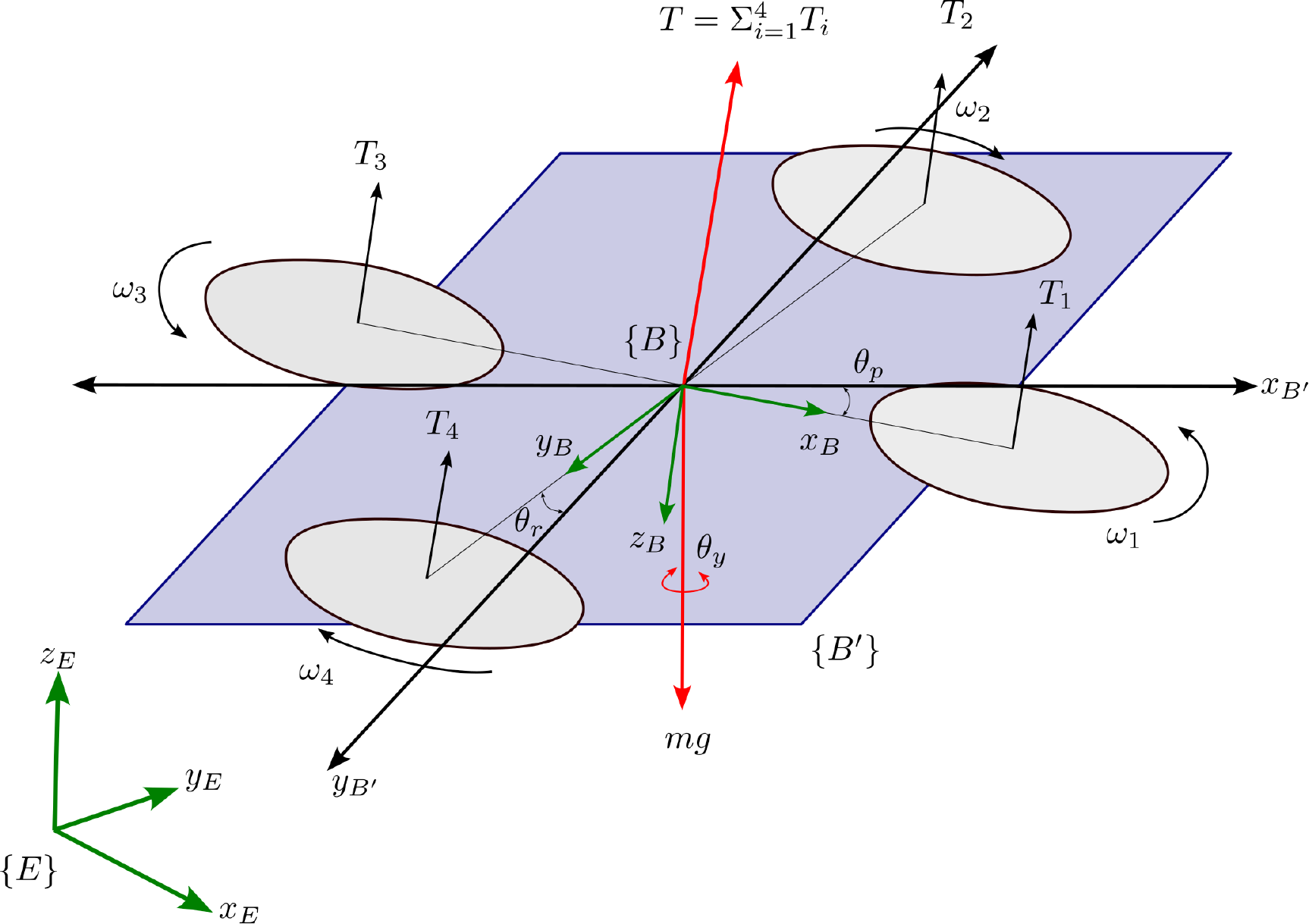}
    \caption{Quadcopter reference frames setup}
    \label{Fig: Quad Dynamics}
\end{figure}

Figure \ref{Fig: Quad Dynamics} shows the standard definitions \cite{Robotics_Vision_Control} of the reference frames used in this simulation. $\{E\}$ is the \textit{Earth reference frame} having its $z$-axis facing upward in the space. The body coordinate frame, $\{B\}$ is attached to the center of the quadcopter having its $z$-axis facing downwards in the space. $\{B^{\prime}\}$ reference frame is attached to the body having its origin same as $\{B\}$ and $x^{B^{\prime}}y^{B^{\prime}}$ plane is parallel to $x^{E}y^{E}$. 

The angles $\theta_{p}$ (pitch), $\theta_{r}$ (roll), and $\theta_{y}$ (yaw) represents the rotation of the body along $x$, $y$, and $z$-axis of the reference frame respectively. As $x^{B^{\prime}}y^{B^{\prime}}$ plane is parallel to $x^{E}y^{E}$ thus $\{B^{\prime}\}$ can only rotate along the z-axis with angle $\theta_{p}$. The change in orientation between $\{B^{\prime}\}$ and \{E\} is given by a rotation matrix
\begin{equation*}
\mathbf{R}_{B^{\prime}}^{E}=\begin{bmatrix}
c \theta_{y} & s \theta_{y} & 0 \\
-s \theta_{y} & c \theta_{y} & 0 \\
0 & 0 & 1
\end{bmatrix}
\end{equation*}
where $ c(.) \coloneqq \cos{(.)} $ and $ s(.) \coloneqq \sin{(.)}$

\subsection{Quadcopter Dynamics}\label{subsec:qd}
The upward thrust along the $-z^{B}$ axis is given by 
\begin{equation*}
    T_{i} = b\bar{\omega}_{i}^2
\end{equation*}
where $\omega_{i}$ is the angular velocity of the rotor and $b>0$ is a lift constant that depends on the air density, number of blades, cube of rotor blade radius, and chord length of the blade.
According to Newton's second law of motion, the translation dynamics of the quadcopter in $\{E\}$ is given by 
\begin{equation}
m\dot{\mathbf{v}}^{E}=\begin{bmatrix}
0 & 0 & m g
\end{bmatrix}^{T}-\mathbf{R}_{B}^{E}\begin{bmatrix}
0 & 0 & T
\end{bmatrix}^{T} - B\mathbf{v}\label{eq:trans_dynamics}
\end{equation}
where $\mathbf{v}$ is the velocity of quadcopter in \{E\} given by $\mathbf{v} =\begin{bmatrix}  {v_{x}}^{E} & {v_{y}}^{E} & {v_{z}}^{E} \end{bmatrix}^{T} \in \mathbb{R}^{3}$ , $B$ is aerodynamics friction and  $T = \Sigma T_{i}$ is total upward thrust generated by the rotors in $-z^{B}$ direction, $g$ is the gravitational acceleration, and $m$ is the mass of quadcopter. In \eqref{eq:trans_dynamics}, first term represents the gravitational force acting in $-z^{E}$ direction, second term represents the total upward thrust generated in  $-z^{B}$ direction in $\{B\}$ and rotated in $\{E\}$.

The rotation in each axis is obtained by varying pairwise differences in rotor thrust. Torque about $x$ and $y$ axes is given by
\begin{equation*}
\tau_{x}=d T_{4}-d T_{2} = db(\bar{\omega}_{4}^{2}-\bar{\omega}_{2}^{2})
\end{equation*}
\begin{equation*}
    \tau_{y}=d T_{1}-d T_{3} = db(\bar{\omega}_{1}^{2}-\bar{\omega}_{3}^{2})
\end{equation*}
The torque applied to each motor is opposed by aerodynamic drag and exerts a reaction torque on the frame which is given by $Q_{i}=k\bar{\omega}_{i}$ where $k$ depends on the same factors as $b$.
The total reaction torque about $z$-axis is given by 
\begin{equation}
\tau_{z} =Q_{1}-Q_{2}+Q_{3}-Q_{4}=k\left(\bar{\omega}_{1}^{2}+\bar{\omega}_{3}^{2}-\bar{\omega}_{2}^{2}-\bar{\omega}_{4}^{2}\right)\label{eq:react_torque_z}
\end{equation}

The total torque applied to the quadcopter body is  $ \mathbf{\Gamma} =\begin{bmatrix}  {\tau}_{x} & {\tau}_{y} & {\tau}_{z} \end{bmatrix}$. By Euler's equation of motion rotational acceleration is given is by 
\begin{equation}
J\boldsymbol{\Dot{\omega}}=-\boldsymbol{\omega} \boldsymbol{\times}J \boldsymbol{\omega}+\mathbf{\Gamma}
\end{equation}
where J is the $3\times3$ inertia matrix of the quadcopter and $\omega$ is a angular velocity vector of the quadcopter body frame.
Overall quadcopter motion equation \cite{Robotics_Vision_Control} is obtained by integrating equations \eqref{eq:trans_dynamics} and \eqref{eq:react_torque_z} which is given by 
\begin{align}
\begin{bmatrix}
\mathbf{T} &
\mathbf{\Gamma}
\end{bmatrix}^T &=A\begin{bmatrix}
\bar{\omega}_{1}^{2} &
\bar{\omega}_{2}^{2} &
\bar{\omega}_{3}^{2} &
\bar{\omega}_{4}^{2}
\end{bmatrix}^T
\end{align}

\noindent where $A =\begin{bmatrix}
-b & -b & -b & -b \\
0 & -d b & 0 & d b \\
d b & 0 & -d b & 0 \\
k & -k & k & -k
\end{bmatrix}$ is a constant invertible matrix since $b,k,d > 0$.

\subsection{Quadcopter Dynamics as Double Integrators}\label{subsec:di}


The overall motion in the plane $x^{B^{\prime}}y^{B^{\prime}}$ is obtained by pitching and rolling the quadcopter by an angle $\theta_{p}$ and $\theta_{r}$. The total force acting on a quadcopter is given by
\begin{equation}
\mathbf{f}^{B^{\prime}}=\mathbf{R}{x}\left(\theta{r}\right) \mathbf{R}{y}\left(\theta{p}\right)\begin{bmatrix}
0 & 0 & T
\end{bmatrix}^{T}\label{eq:total_force}
\end{equation}
where,
\begin{align*}
\mathbf{R}{x}\left(\theta{r}\right)=\begin{bmatrix}
1 & 0 & 0 \\
0 & c \theta_{r} & s \theta_{r} \\
0 & -s \theta_{r} & c \theta_{r}
\end{bmatrix},
~
    \mathbf{R}{y}\left(\theta{p}\right)=\begin{bmatrix}
c \theta_{p} & 0 & s \theta_{p} \\
0 & 1 & 0 \\
-s \theta_{p} & 0 & c \theta_{p}
\end{bmatrix}
\end{align*}
are the Rotation matrix about $x$ and $y$ axes respectively.\\
Simplifying \eqref{eq:total_force} we get $\mathbf{f}^{B^{\prime}}$ as 
\begin{equation*}
\mathbf{f}^{B^{\prime}}=\begin{bmatrix}
T\sin{\theta_{p}} & T\sin{\theta_{r}}\cos{\theta_{p}} & T\cos{\theta_{r}}\cos{\theta_{p}}
\end{bmatrix}^{T}
\end{equation*}
for small $\theta_{p}$ and $\theta_{r}$ the above equation can be approximated by 
\begin{equation*}
\mathbf{f}^{B^{\prime}} \approx \begin{bmatrix}
T\theta_{p} & T\theta_{r} & T
\end{bmatrix}^{T}
\end{equation*} 
i.e $f_{x}^{B^{\prime}} \approx T\theta_{p}$ and $f_{y}^{B^{\prime}} \approx T\theta_{r}$ \\
According to Newton's Second law of motion $F=ma$, we can write
\begin{equation*}
 ma_{x}^{B^{\prime}} = T\theta_{p}
\end{equation*}
\begin{equation*}
    \theta_{p} = \dfrac{m}{T}a_{x}^{B^{\prime}},~~\theta_{r} = \dfrac{m}{T}a_{y}^{B^{\prime}}
\end{equation*}
where $a_{x}^{B^{\prime}}$ is  $a_{y}^{B^{\prime}}$ is acceleration of quadcopter in $\{B^{\prime}\}$ frame.\\
As the controller will receive the position and velocity in $\{E\}$ frame so it will compute the acceleration in $\{E\}$ frame which needs to be rotated to the $\{B^{\prime}\}$ frame.
\begin{equation*}
    \mathbf{f}^{B^{\prime}} = \mathbf{R}_{E}^{B^{\prime}}\mathbf{f^{E}}
\end{equation*}
\begin{equation*}
    m\mathbf{a}^{B^{\prime}} = \mathbf{R}_{E}^{B^{\prime}}m\mathbf{a^{E}}
\end{equation*}
\begin{equation*}
    \mathbf{a}^{B^{\prime}} = \mathbf{R}_{E}^{B^{\prime}}\mathbf{a^{E}}
\end{equation*}

This equation shows the motion of the quadcopter in $x^{B^{\prime}}y^{B^{\prime}}$ plane.

\begin{rem}
 While the discussion above describes the motion of the quadcopter in $x^{B^{\prime}}y^{B^{\prime}}$ plane as a double integrator, motion along $-z^{B^\prime}$ is directly dependent on the thrust generated by the rotors in \{B\} frame given by $T = \Sigma T_{i}$, which is naturally a double integrator.
\end{rem}

\subsection{Preliminaries of Distributed Control}
A homogeneous multi-agent system is a collection of $n$ agents with identical dynamics communicating with each other over a communication graph $\mathcal{G}=(\mathcal{V}, \mathcal{E})$, where the set of vertices $\mathcal{V}=\{\alpha_i,~i=1,...,n\}$ represents the agents and the edges $(\alpha_i,\alpha_j)\in\mathcal{E}$ denote the communication from agent $\alpha_i$ to $\alpha_j$. The communication graph $\mathcal{G}$ is \textit{undirected} if $(\alpha_i,\alpha_j)\in\mathcal{E}\implies(\alpha_j,\alpha_i)\in\mathcal{E}$
otherwise it is \textit{directed}. A \textit{path} in a graph is a sequence of edges. An undirected graph is said to be \textit{connected} if it has a path from each agent to every other agent. A directed graph is said to contain a \textit{directed spanning tree} if there is an agent $\alpha_r$ (called \textit{root}) such that there is a directed path from $\alpha_r$ to every other agent. Set of neighbours for each agent is given by $\mathcal{N}_{i} := \{j \in \mathcal{V} : (\alpha_i,\alpha_j) \in \mathcal{E} \}$. The Laplacian matrix $\mathcal{L}$ of a graph $\mathcal{G}$ is given by $\mathcal{L}_{n}=\left[l{i j}\right] \in \mathbb{R}^{n \times n} ; l_{i j}=-a_{i j}, i \neq j, l_{i i}=\sum_{j=1, j \neq i}^{n} a_{i j} $, where $a_{ij}$ is the \textit{weight} of the edge $(\alpha_i, \alpha_j)$.
Some control algorithms use leader-follower configuration of the multi-agent systems in which, the set of agents is divided into a set of leader $\mathbf{L}$ and a set of followers $\mathbf{F}$.

In this paper, agents are assumed to have double-integrator dynamics, given by 
\begin{equation}
\ddot{\mathbf{x}}_i^E(t)=\mathbf{f}_i^E(t)\label{eq:agent}
\end{equation}
\noindent where $\mathbf{x}_i^E=\begin{bmatrix}\mathbf{p}^E_i\\\mathbf{v}^E_i\end{bmatrix}\in\mathbb{R}^{2d}$ is the state vector of $\alpha_i$ and $\mathbf{f}^E_i\in\mathbb{R}^d$ denotes the input to $\alpha_i$. The vectors $\mathbf{p}^E_i$ and $\mathbf{v}^E_i$ denote the position and velocity of $\alpha_i$, respectively, in $\{E\}$ frame. 


\section{MASCOT: Structure and Features}\label{sec:mascot}

The testbed presented in this paper "MASCOT" is based on open source tools, ROS and Gazebo. For the demonstration purpose, we have adapted the AR drone Gazebo plugin \cite{Tum} to represent quadcopter agents. Using the discussion presented in Section \ref{sec:prelim}, the control plugin of the AR drone is modified such that the agents can be treated as double integrators.

\subsection{Tools Used}\label{subsec:tools}
\subsubsection{Robot Operating System (ROS)}
\begin{figure}[!ht]
\centering
    \includegraphics[scale=0.15]{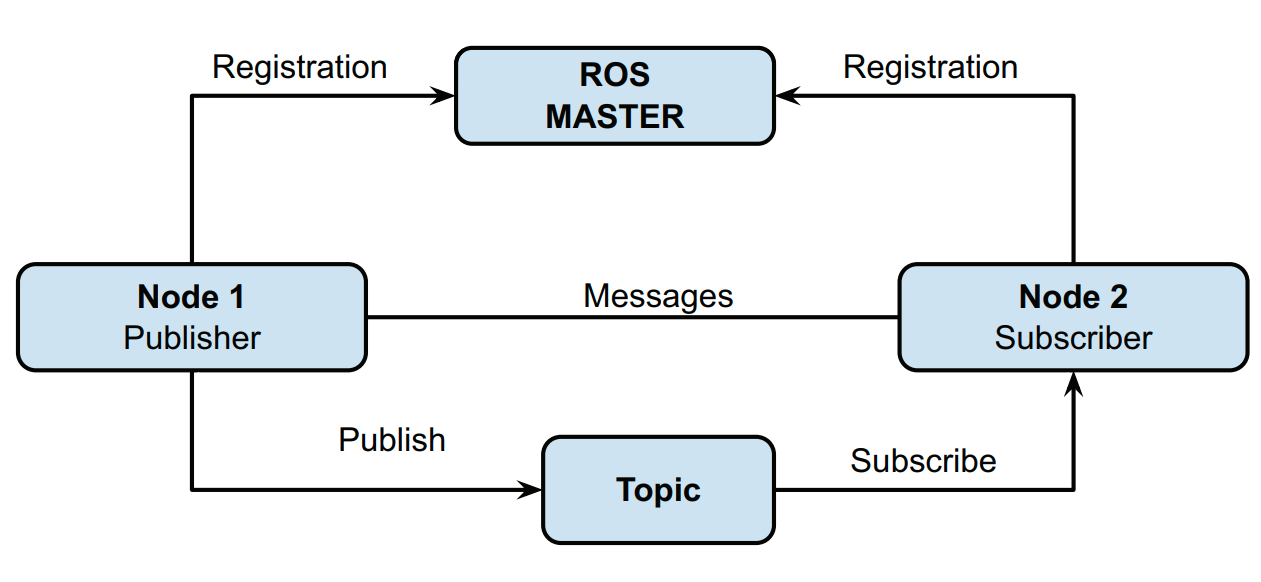} 
    \caption{ROS Architecture}
    \label{Fig:ROS Architecture}
\end{figure}

ROS is an open-source framework that helps researchers and developers build and reuse code between robotics applications. ROS provides a distributive architecture that eliminates the problem of a single node handling all the tasks. ROS supports low-level drivers and programming in various languages such as Python, C++, Java, and Lisp. This makes it convenient to directly deploy the codes written for simulation on the physical systems. Visualization and debugging tools such as Rviz and rqt make it very easy to keep track of the process\cite{ros}. The basic architecture of ROS is shown in Fig. \ref{Fig:ROS Architecture}.

\subsubsection{Gazebo}
The gazebo is a 3D simulation platform that provides robots, sensors, and environment models which are required for robot development. It uses Open Dynamics Engine (ODE) for real-time simulations and graphics. Gazebo supports a wide range of sensors such as LRF,2D/3D camera, Depth Camera, and Force-Torque Sensor. APIs provided in Gazebo enable users to create robots, sensors, and environment control as a plug-in.

\subsubsection{Quadcopter Simulation Package}
The quadcopter model used in MASCOT is based on the tum-simulator AR Parrot Drone Gazebo simulation package \cite{Tum} developed by the Computer Vision Group at the Technical University of Munich. The lower level controller is modified and tuned so that the quadcopters can be treated as a double integrator system for control algorithm implementation, and added topics and controls required for the implemented model. A general control script is written for the implementation of the different use cases discussed in Section \ref{sec: Use Cases};

\subsection{Control Block}

\begin{figure}[!ht]
\centering
    \includegraphics[scale=0.40]{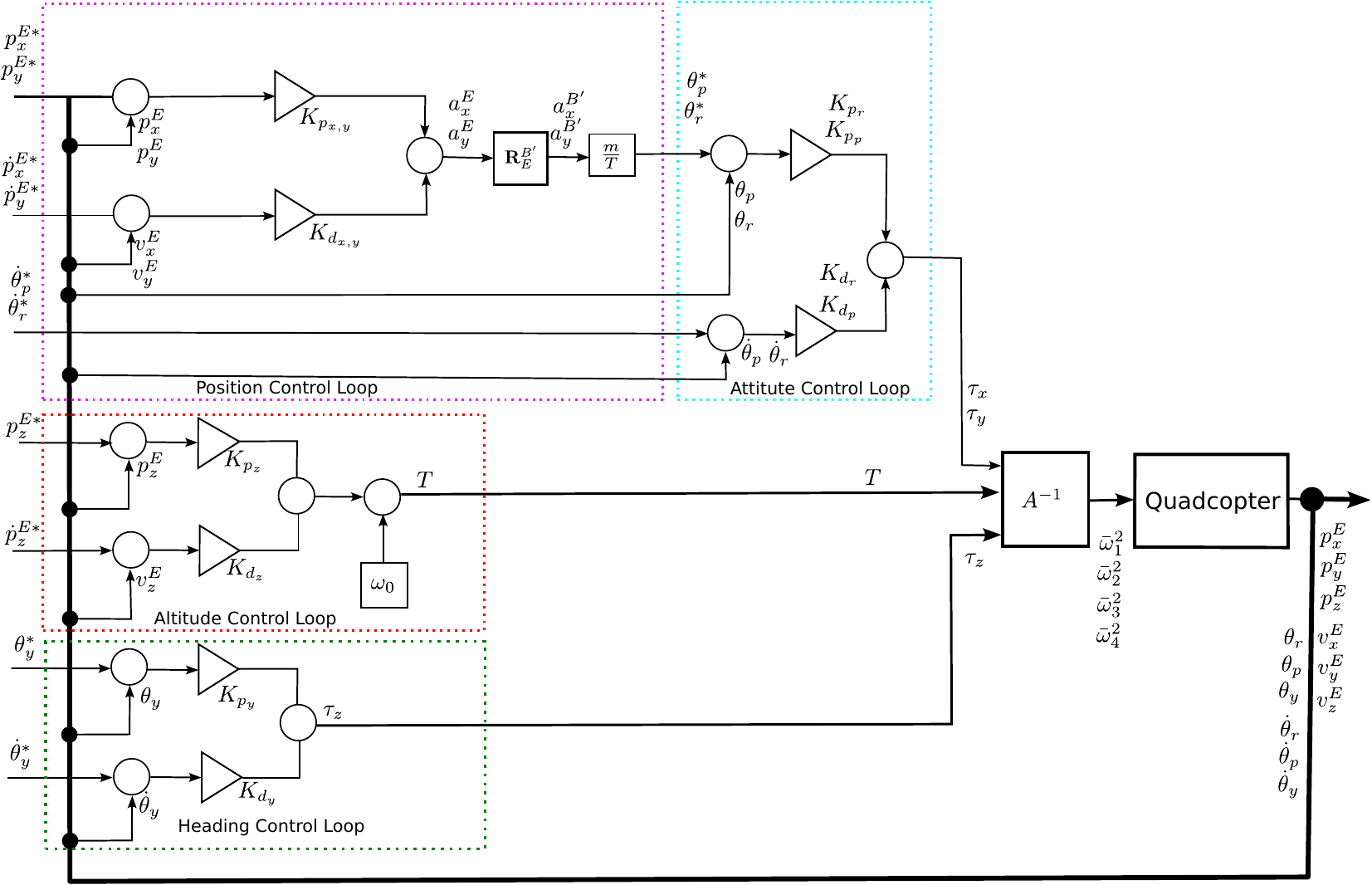} 
    \caption{Control Block diagram}
    \label{Fig: Control Block Diagram}
\end{figure}
Figure \ref{Fig: Control Block Diagram} shows the Control block diagram of the system.
The low-level control of the quadcopter is achieved by four independent controllers given below:
\begin{itemize}
    \item Heading Control
    \item Altitude Control
    \item Pitch and Roll angle control
    \item Position Control
\end{itemize}
\subsubsection{Altitude Control}
The Altitude of the quadcopter is achieved by the total Thrust $T$ generated by all the rotors. This thrust is controlled by the following proportional-derivative controller.
\begin{equation*}
T=K_{p_{z}}\left(p_{z}^{*}-p_{z}\right)+K_{d_{z}}\left(\dot{p}_{z}^{*}-\dot{p}{z}\right)+\omega_{0}
\end{equation*}
where $p_{z}$ is the current altitude, $p_{z}^{*}$ is the desired altitude, $K_{p}$, $K_{d}$ are proportional, derivative gains respectively and $\omega_{0}$ is the rotor speed bias which generates a trust to counter the weight of quadcopter

\subsubsection{Heading Control}
The heading is a yaw angle $\theta_{y}$ which represents the angle between $x^{B}$ and $x^{E}$ about $z^{E}$axis. To control this angle we generate a required torque $\tau_{z}$ which is given by the following proportional-derivative controller.
\begin{equation*}
\tau_{z}=K_{p_{y}}\left(\theta_{y}^{*}-\theta_{y}\right)+K_{d_{y}}\left(\dot{\theta}_{y}^{*}-\dot{\theta}_{y}\right)
\end{equation*}
where $\theta_{y}$ is a current yaw angle, $\theta_{y}^{*}$ is the desired yaw angle and $K_{p}$,$K_{d}$ are proportional,derivative gains respectively.
\subsubsection{Controlling $\theta_{r}$ and $\theta_{p}$}
The roll angle $\theta_{r}$ represents the angle between $y^{B^{\prime}}$ and $y^{B}$ about $x^{B^{\prime}}$ axis.To control this angle we generate a required torque $\tau_{r}$ which is given by the following proportional-derivative controller.
\begin{equation*}
\tau_{x}=K_{p_{r}}\left(\theta_{r}^{*}-\theta_{r}\right)+K_{d_{r}}\left(\dot{\theta}_{r}^{*}-\dot{\theta}_{r}\right)
\end{equation*}
where $\theta_{r}$ is a current roll angle,$\theta_{r}^{*}$ is the desired roll angle and $K_{p_{r}}$, $K_{d_{r}}$ are proportional,derivative gains respectively.

The pitch angle $\theta_{y}$ represents the angle between $x^{B^{\prime}}$ and $x^{B}$ about $y^{B^{\prime}}$ axis.To control this angle we generate a required torque $\tau_{p}$ which is given by the following proportional-derivative controller.
\begin{equation*}
\tau_{y}=K_{p_{p}}\left(\theta_{p}^{*}-\theta_{p}\right)+K_{d_{p}}\left(\dot{\theta}_{p}^{*}-\dot{\theta}_{p}\right)
\end{equation*}
where $\theta_{p}$ is a current pitch angle,$\theta_{p}^{*}$ is the desired pitch angle and $K_{p_{p}}$, $K_{d_{p}}$ are proportional,derivative gains respectively.

\begin{rem} With the change in $\theta_{p}$ and $\theta_{r}$ there is also a reduction in the vertical component of the total thrust T by the factor of the cosine of $\theta_{p}$ and $\theta_{r}$. As per our assumption for a very small angle, this reduction is not much which is rapidly corrected by an Altitude control loop. 
\end{rem}

\subsubsection{Position Control}
The position of the quadcopter in $x^{B^{\prime}}y^{B^{\prime}}$ plane is controlled independently by  the proportional-derivative controller for each axis.
\begin{equation*}\label{x-pos-control}
   a_{x} = K_{p_{x}}\left(p_{x}^{*}-p_{x}\right)+K_{d_{x}}\left(\dot{p}_{x}^{*}-\dot{p}{x}\right) 
\end{equation*}
\begin{equation*}\label{y-pos-control}
   a_{y} = K_{p_{y}}\left(p_{y}^{*}-p_{y}\right)+K_{d_{y}}\left(\dot{p}_{y}^{*}-\dot{p}{y}\right) 
\end{equation*}
\textbf{Note:} Here the $a_{x}$ and $a_{y}$ computed is in \{E\} as feedback position and velocity is also in \{E\} frame.

\subsection{Architecture and Features of MASCOT}
\begin{figure}[!ht]
\centering
    \includegraphics[scale=0.55]{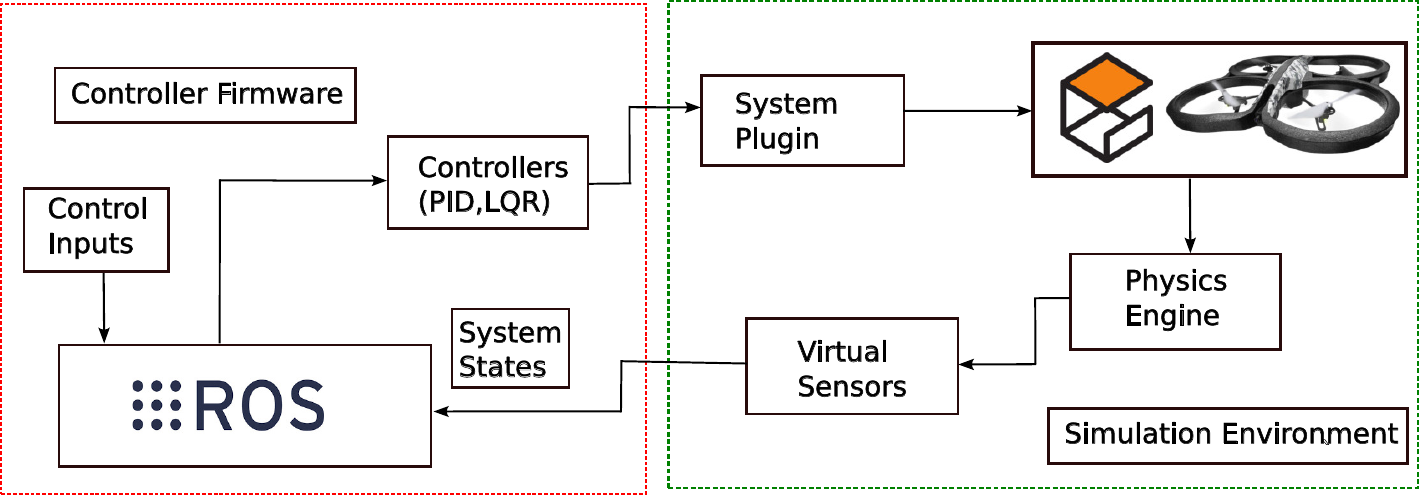} 
    \caption{System Architecture}
    \label{Fig:System Architecture}
\end{figure}

The simulation architecture of MASCOT is shown in Fig. \ref{Fig:System Architecture}.
In this system, the Gazebo simulator simulates the model and its sensors with the system plugin. Gazebo's internal scheduler provides the ROS interfaces. This ROS interface enables a user to manipulate the properties of the simulation environment over ROS, as well as spawn and introspect on the state of models in the environment. ROS works as a middleware that runs the independent controller of individual robots spawned in Gazebo environments, ROS provides the intercommunication channel within the nodes, where all the controller nodes communicate with each other and Gazebo over the TCPROS protocol and exchange the information. The control node is written in python for the high-level control of the quadcopter the low-level control and plugin are developed in C++ which provides the tuned low-level control of the quadcopter. If needed user can modify or tune this low-level control.
\subsection{Configuring the Simulation}
For simulating the multi-agent systems using MASCOT, users can provide configuration parameters and control laws using a plain text configuration file, which is used by the internal programs for the initialization and simulation of the quadrotor.

The configurable parameters required are as follows: 
\begin{itemize}
    \item \textbf{Robot:} Details of the Robots to be simulated
        \begin{itemize}
            \item \textbf{Number:} No. of Agents to be spawned in Gazebo.
            \item \textbf{IntialPosition:} Flag to enable random initial position or specified initial position.
            \item \textbf{Position:} Initial Position of Robot (x,y,z).
        \end{itemize}
    \item \textbf{Control: } Controls laws to be used
        \begin{itemize}
            \item \textbf{Custom-Control: } Flag to Enable Custom Control.
            \item \textbf{Tutorial Examples: } Flag to Enable Tutorial.
            \begin{enumerate}
                \item \textbf{Waypoint Navigation:}
                \begin{itemize}
                    \item \textbf{P-Gain: } Proportial Gain , Default = 1.0
                    \item \textbf{D-Gain: } Differential Gain , Default = 1.0
                \end{itemize}
                \item \textbf{Consensus: } Flag to Enable Consensus Control.
                \begin{itemize}
                    \item \textbf{Leader: } Specify the robot number that to be leader, 0-for leaderless consensus.
                    \item \textbf{Communication Graph: } Flag for communication graph 
                    \item \textbf{L-mat: } Laplacian Matrix.  
                \end{itemize}
                \item \textbf{Min-max Consensus: }Enable Min-max Consensus Control.
                \end{enumerate}
            \end{itemize}
    \item \textbf{Output:}  Configuration of output
        \begin{itemize}
            \item \textbf{Velocity : } Enable it to generate the velocity plot.
            \item \textbf{Position : } Enable it to generate the position plot.
            \item \textbf{Save-plot : } Enable to save the plots.
            \item \textbf{Show-plot : } Enable it to show plots.
            \item \textbf{Save-data : } Enable to save as numpy-array
        \end{itemize}
\end{itemize}

A user needs to specify their control laws by setting the Custom-Control flag (set as default) as plain text using python syntax. The user can use the states of the quadcopter such as position, velocity, acceleration, and orientation in the control algorithm and can publish the acceleration command which is an output of the control algorithm. Few demos are made available as tutorials for the users which are described in Section \ref{sec: Use Cases} below.

Theoretically, there is no limit on the number of agents in MASCOT, however, the simulation performance may be affected based on the number of agents and computer specifications. The effective control of the quadcopters depends on the system specifications as well as the computational complexity of the algorithm being tested.


\section{Examples}\label{sec: Use Cases}

In this section, the simulation results of MASCOT are demonstrated using a few examples of linear as well as non-linear distributed control laws designed for multi-agent systems. A fewer number of agents are used for clarity of figures. All the Algorithms are tested on the Personal Computer with Intel i7 running at 2.8 GHz and 16 GB RAM. The OS used is Ubuntu 20.04.4 with ROS Noetic and Python 3.8.10.

\begin{figure}[!ht]
   \centering
   \subfigure[Initial position of quadcopters at t = 0 s]
   {
       \includegraphics[scale=0.06]{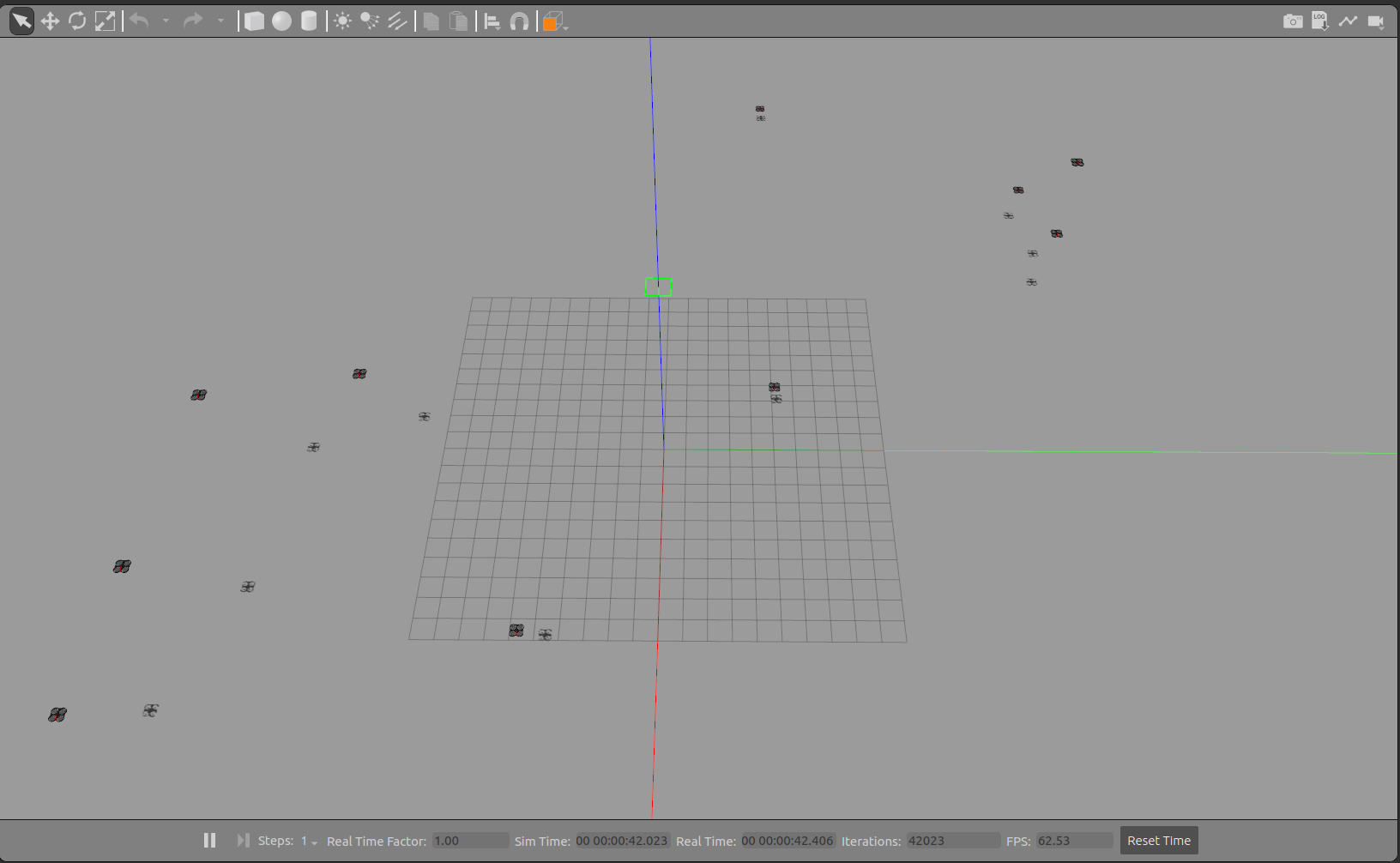}
       \label{fig:Initial position of quadcopters}
   }
   \subfigure[Final position of quadcopters after consensus at t = 80 s]
   {
       \includegraphics[scale=0.06]{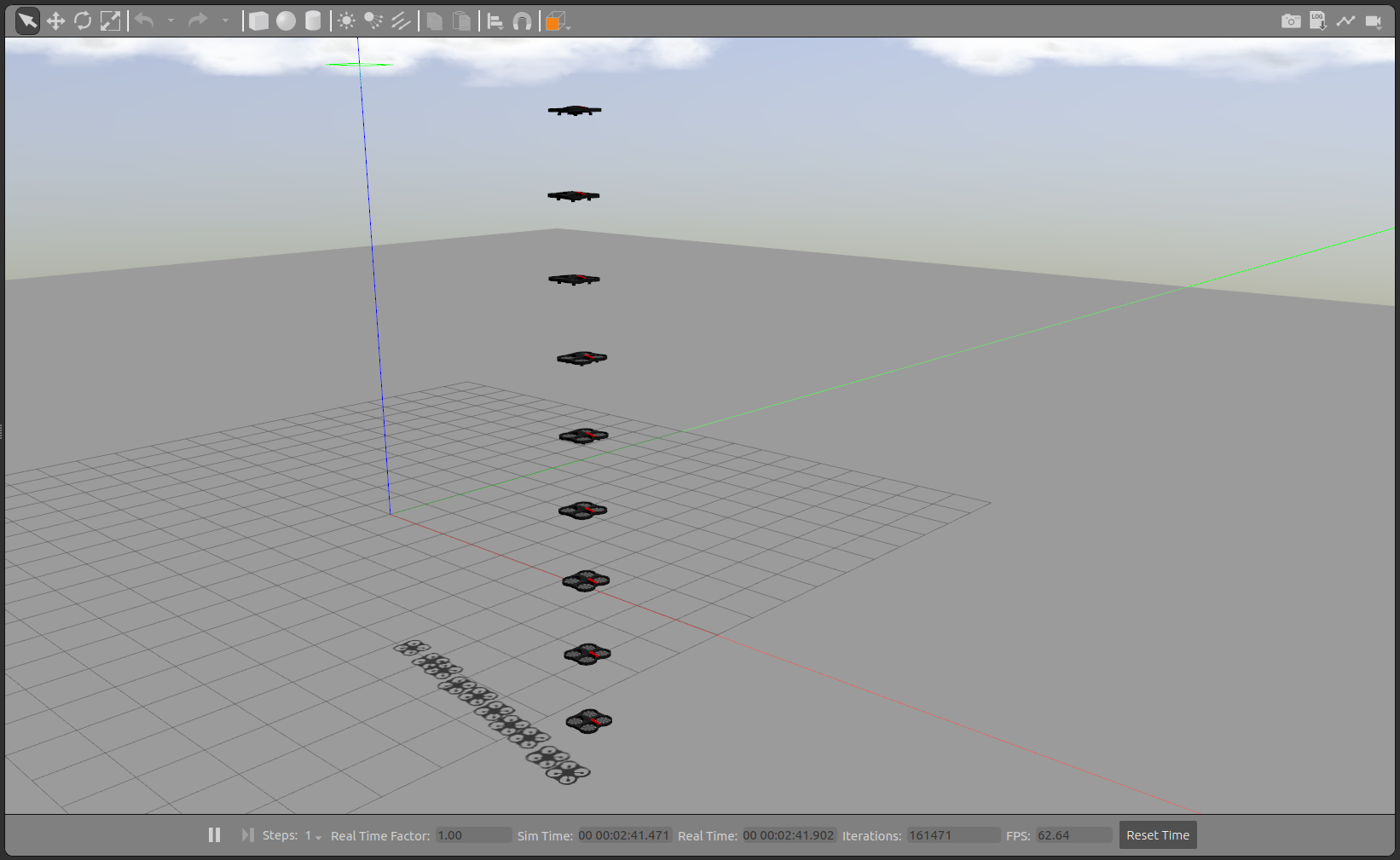}
       \label{fig:Final position of quadcopters}
   }
   \caption{Visualization in Gazebo}
   \label{Gazebo_visualization}

\end{figure}

\subsection{Waypoint Navigation}
\begin{figure}[!ht]
    \centering
    \subfigure[Position in X-axis]
    {
        \includegraphics[scale=0.175]{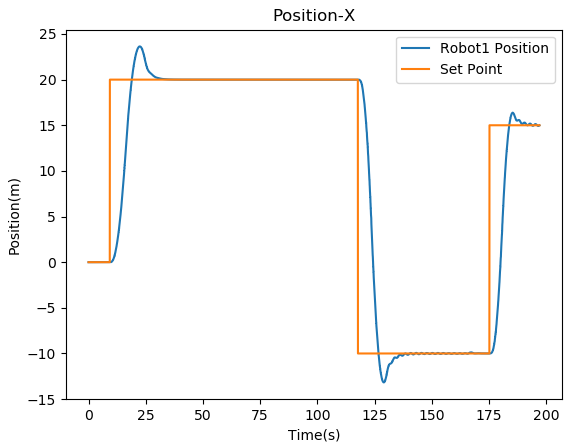}
        \label{fig:Position in X-axis}
    }
    \subfigure[Position in Y-axis]
    {
        \includegraphics[scale=0.175]{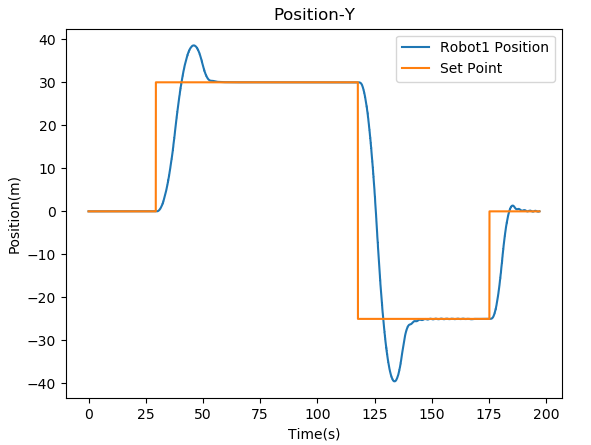}
        \label{fig:Position in Y-axis}
    }
    \caption{Quadcopter Position Plots.}
    \label{fig:Quadcopter Position Plots}
\end{figure}

The quadcopter is initialized at a $(0,0,1)$ position and the different waypoints are given in the $x^Ey^E$ plane. From Figure \ref{fig:Quadcopter Position Plots}, it can be inferred that the quadcopter achieves its goal position with a very small overshoot, which depends on the parameters of the controller.

\subsection{Consensus Algorithms}
We demonstrate two consensus algorithms that have linear control laws  \cite{Distributed_Consensus_book}: 1) for leaderless asymptotic consensus, and 2)  leader-follower configuration i.e. consensus tracking
In a leaderless case, four agent system is considered which are initialized at random positions, while in leader-follower configuration a ten-agent system is considered with $\alpha_4\in\textbf{L}$. 

\begin{figure}[!ht]
    \subfigure[Leaderless Control plot]
    {
        \includegraphics[scale=0.175]{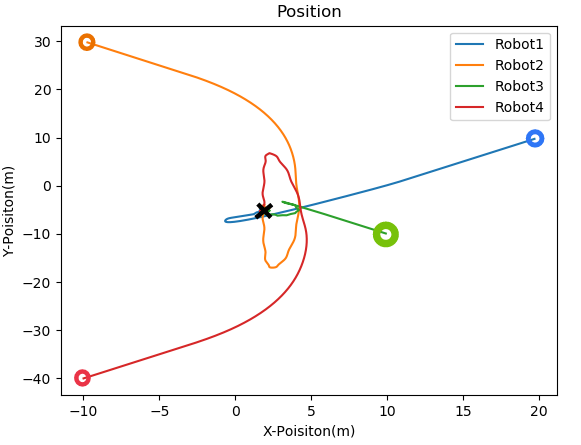}
        \label{fig:Position plot distributed}
    }
    \subfigure[Leader Follower plot]
    {
        \includegraphics[scale=0.175]{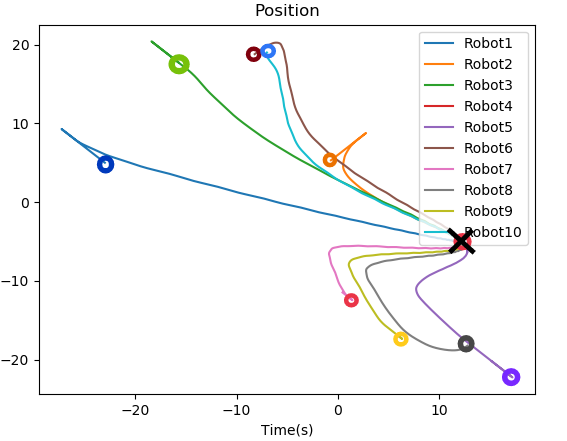}
        \label{fig:Position plot leader}
    }
    \caption{ Consensus Control Position Plots.}
    \label{fig:Quadcopter Consensus Control Plots }
\end{figure}

\begin{figure}[!ht]
    \subfigure[Position in X-axis]
    {
        \includegraphics[scale=0.175]{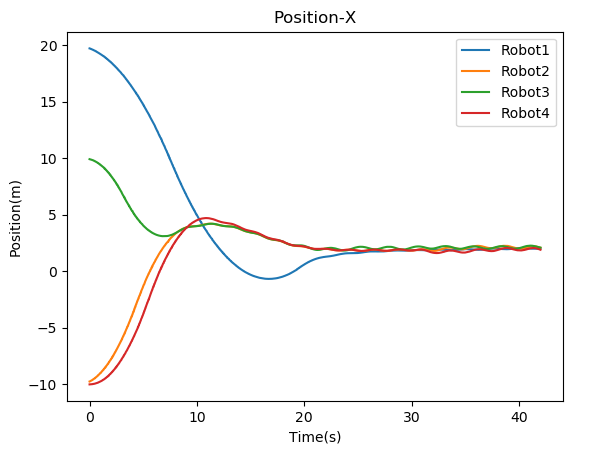}
        \label{fig:Position in X-axis leaderless}
    }
    \subfigure[Position in Y-axis]
    {
        \includegraphics[scale=0.175]{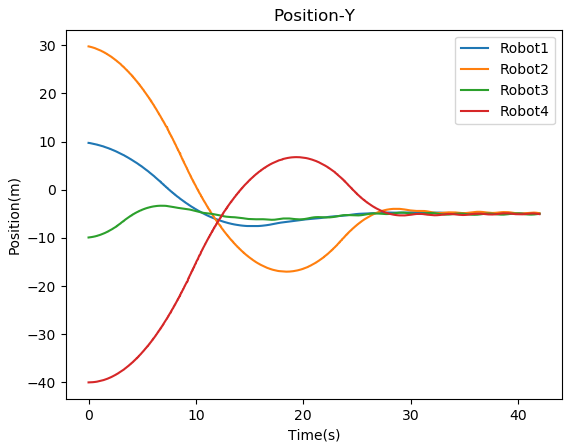}
        \label{fig:Position in Y-axis leaderless}
    }
    \caption{Quadcopter Leaderless Control Plots.}
    \label{fig:Quadcopter Consensus leaderless Control Plots}
\end{figure}

\begin{figure}[!ht]
    \subfigure[Position in X-axis]
    {
        \includegraphics[scale=0.175]{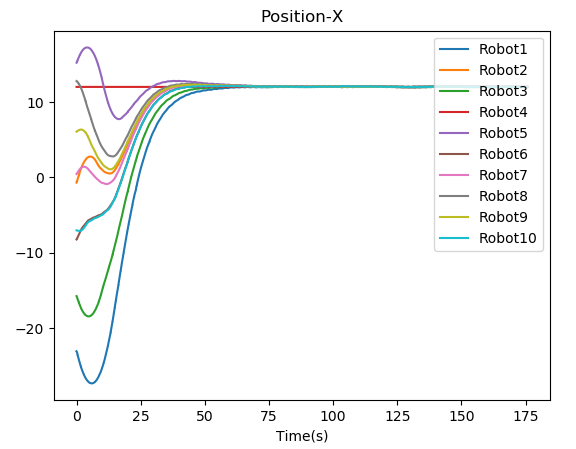}
        \label{fig:Position in X-axis leader}
    }
    \subfigure[Position in Y-axis]
    {
        \includegraphics[scale=0.175]{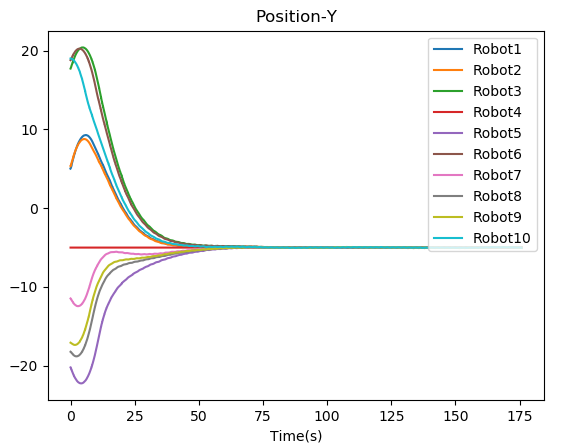}
        \label{fig:Position in Y-axis leader}
    }
    \caption{Quadcopter Leader Consensus Control Plots.}
    \label{fig:Quadcopter Consensus leader Control Plots}
\end{figure}

The control algorithms used is as follows:
\begin{equation}
\textbf{f}_{i}^{E}=\left\{\begin{array}{l}
\sum_{j = 1}^{n} a_{ij}\left(\mathbf{p}{j}^{E}-\mathbf{p}{i}^{E}\right)-\beta \textbf{v}_{i}^{E} \text { if } \alpha_{i} \in \mathbf{F} \\
\\
0 \text { if }  \alpha_{i} \in \mathbf{L}
\end{array}\right.
\end{equation}
where $\beta$ is a positive constant. Note that, in a leaderless case, all the agents are assumed to be followers.
All the quadcopters run their controllers simultaneously and plugin under their respective namespace. Figure \ref{Gazebo_visualization} shows the visualization of the convergence of quadcopters in the Gazebo simulator. For the leaderless case, Fig. \ref{fig:Position plot distributed} shows the trajectory followed by all the quadcopters to reach the consensus point. Fig. \ref{fig:Quadcopter Consensus leaderless Control Plots} shows the convergence of the position of the quadcopter in the $x^E$ and $y^E$ axis.
 For leader-follower configuration Fig. \ref{fig:Position plot leader} shows the trajectory followed by the follower and Fig. \ref{fig:Quadcopter Consensus leader Control Plots} show the convergence of the position of the Quadcopter in the $x^E$ and $y^E$-axis.

\subsection{Min-max time Consensus Control}
To demonstrate the simulations of non-linear distributed control laws, a min-max time consensus tracking algorithm proposed in \cite{mulla2017min} is used. This control algorithm extracts a directed spanning tree from the communication graph between the agents and treats each edge as a leader-follower pair. The root of the spanning tree gives the reference trajectory for consensus tracking. The agents are assumed to have bounded inputs. Using the difference of states between the leader ($\alpha_p$) and the follower ($\alpha_c$) in each edge, a distributed bang-bang control law is given for the following agents:
\begin{equation}
    \mathbf{f}^E_c=\beta_c\text{sign}(2(\beta_c-\beta_p)(\mathbf{p}_c-\mathbf{p}_p)+(\mathbf{v}_c-\mathbf{v}_p)^2\text{sign}(\mathbf{v}_c-\mathbf{v}_p))
\end{equation} where $\beta_p$ and $\beta_c$ are the input bounds of $\alpha_p$ and $\alpha_c$ respectively. 

\begin{figure}[!ht]
    \subfigure[Position in X-axis]
    {
        \includegraphics[scale=0.175]{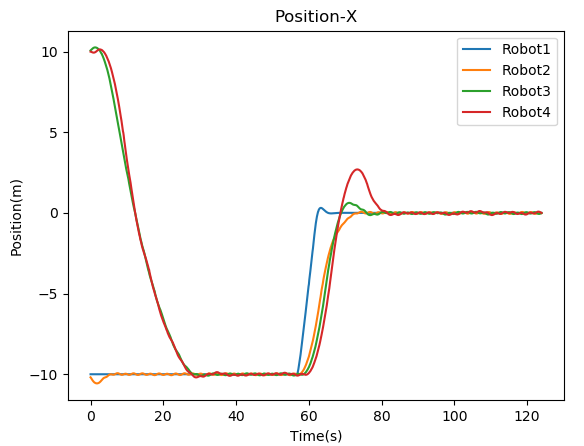}
        \label{fig:Position in X-axis minmax}
    }
    \subfigure[Position in Y-axis]
    {
        \includegraphics[scale=0.175]{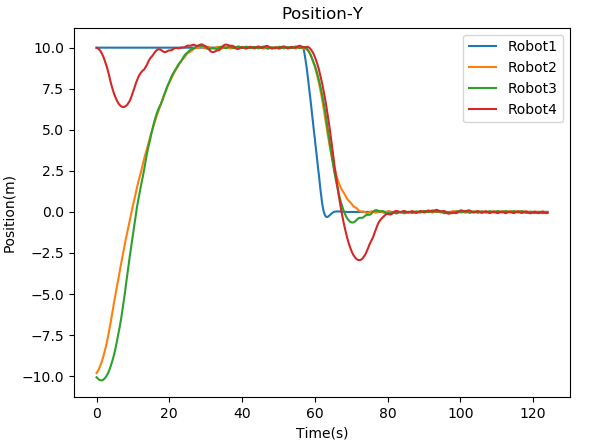}
        \label{fig:Position in Y-axis minmax}
    }
    \caption{Quadcopter minmax Consensus Control Plots.}
    \label{fig:Quadcopter Consensus minmax Control Plots}
\end{figure}

For simulation, a four-agent system is used with $\alpha_1$ as the root agent. Figure \ref{fig:Quadcopter Consensus minmax Control Plots} shows the test result for this control. 

\section{Conclusion and Future Work} \label{sec:conc}

MASCOT is a simulation testbed developed for testing distributed control laws designed for multiagent systems with double integrator agents on quadcopter models, various linear, as well as non-linear control laws, can be tested and few of them are demonstrated in this article. This shows the integrity of the developed testbed and also a guide to using it. Further, this testbed can be expanded for other homogeneous and heterogeneous multiagent systems. As the system is open-source and expandable different interface modules can be developed for MATLAB and Simulink which can help researchers to test their algorithms more efficiently.


\bibliographystyle{IEEEtran}
\bibliography{IEEEabrv,references}


\end{document}